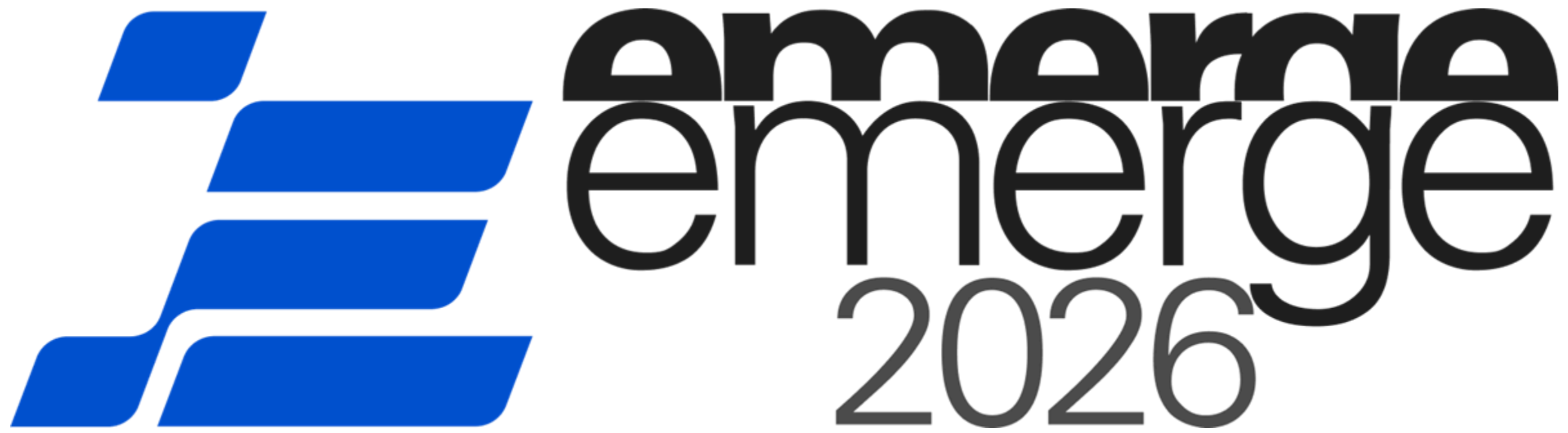

# Paper title

## Participatory Design for Assistive Mobility in Indian Homes, Grounded in Lived Experience

Jyoti Rautela[a], Abinash Kumar Swain[a], Madhan kumar Vasudevan[a]

[a]Indian Institute of Technology Roorkee

jyoti_r@design.iitr.ac.in , abinash.swain@me.iitr.ac.in , madhan.vasudevan@design.iitr.ac.in

**Abstract**: Assistive mobility devices support independence for people with lower-limb disabilities, yet many are designed and evaluated in clinical or controlled environments. In Indian households, narrow spaces and dense furniture often make assistive devices difficult to use indoors, leading people to rely on improvised movements or support from family members and caregivers. In this work, we explore domestic mobility through a participatory and co-speculative design approach, focusing on how people with lower-limb disabilities navigate and maneuver within their homes. We conducted a series of semi-structured interviews and bespoke booklet-based participatory workshops with 22 participants with lower-limb impairments. To support reflection and discussion, we designed bilingual bespoke booklets grounded in domestic design frictions, using images and scenarios to encourage storytelling, critique, and speculation. Our findings reveal mobility challenges that differ significantly from those typically observed in clinical contexts. Rather than yielding a fixed set of design solutions, the study contributes situated insights into domestic mobility frictions, participant articulation, and the limits of speculative participation in this context.

# 1. Introduction

Most existing assistive mobility technologies are designed to accommodate public environments, such as wide corridors, transit stations, ramps, and spaces with guardrails and level flooring. In contrast, their suitability for domestic interiors remains limited. Within homes, particularly in dense or furniture-filled layouts, commonly used assistive devices-especially wheelchairs-are often too large or rigid to support effective indoor navigation. As a result, users frequently experience difficulty manoeuvring through narrow passages, accessing bathrooms or kitchens, and moving between rooms, which increases dependence on family members and caregivers and can negatively affect dignity and autonomy (Imrie, 2004; Phillips & Zhao, 1993).

These limitations contribute to high rates of assistive technology abandonment, a phenomenon widely documented in prior research. Devices designed primarily for outdoor or institutional use often fail to integrate into everyday domestic routines, leading users to discontinue use once they return home (Phillips & Zhao, 1993; Scherer, 2005). For individuals living in single - or multi-storey houses, the need to transition between a heavy, outdoor-oriented mobility device and informal or assisted indoor strategies further increases this burden, making everyday life cumbersome.

Despite advances in inclusive infrastructure and policy-driven improvements in public accessibility, domestic living conditions-particularly in Indian middle-class households-remain underexplored in assistive technology research. Homes are typically densely furnished and spatially constrained, leaving limited clearance for movement with mobility aids. These spatial conditions restrict independent movement and often add reliance on family members for basic daily activities, including transfers, bathing, and food preparation.

At the same time, patterns of work and domestic life are shifting. Alongside individuals who commute for employment, many people-including people with disabilities-work from home, manage household responsibilities, or engage in home-based livelihoods. In both scenarios, access to and agency within the home environment are central to participation and well-being. For people with disabilities to be recognised as equal members of the household, domestic spaces must support autonomy rather than merely accommodate presence (Imrie, 2004). Assistive mobility within the home therefore matters not only for functional movement, but also for social participation, dignity, and a sense of belonging.

Existing assistive device research has largely prioritised technical feasibility, and clinical performance, often with limited involvement of users in early stages of design and decision-making. As a result, devices are frequently developed from the perspectives of engineers, clinicians, or designers, with user input occurring late in the process or being reduced to validation rather than co-creation (Scherer, 2005; Federici et al., 2016). This disconnect has been identified as a key factor contributing to dissatisfaction and subsequent abandonment of assistive technologies (Phillips & Zhao, 1993).

Although substantial work has examined user needs and reasons for abandonment, there remains a lack of design-oriented frameworks that help researchers and practitioners understand how assistive technologies are negotiated within everyday domestic life. In particular, domestic challenges are often treated as secondary to public accessibility, leaving a gap in knowledge about how devices intersect with household routines, spatial constraints, and social dynamics.

Human-centred design frameworks have sought to increase user involvement in product development, yet users often remain positioned as respondents rather than active contributors. Participatory design extends this framework by engaging users as collaborators throughout the design process, recognising their lived experience as a form of expertise (Sanders & Stappers, 2008; Light et al., 2017). Speculative design further complements this approach by using imagined scenarios and fictional artefacts to question existing assumptions and explore alternative possibilities beyond current technological constraints (Dunne & Raby, 2013).

In the context of assistive mobility and lower-limb disability, participatory and speculative approaches are particularly relevant because they enable dialogue that may not emerge through interviews alone. By prompting participants to engage with imagined futures, visual artefacts, and everyday scenarios, these methods can support reflection on domestic challenges, aspirations, and unarticulated needs. When applied carefully, such approaches can help participants express experiences that are otherwise normalised or difficult to verbalise, providing designers and researchers with richer insight into everyday life with assistive technologies.

This paper presents a scenario-based exploratory study involving twenty two participants with lower-limb disabilities, focusing on their domestic mobility challenges and interactions with existing assistive technologies. Through the use of participatory and co-speculative bespoke booklets, the study examines how dialogue around domestic living conditions, assistive device use, and everyday adaptations is elicited.

Rather than proposing new assistive devices, this work contributes a methodological perspective, demonstrating how participatory speculative artefacts can surface otherwise overlooked aspects of domestic life and highlight limits in current assistive technology practices. The findings offer insights for researchers and designers seeking to better understand the relationship between assistive mobility, domestic space, and abandonment, and suggest directions for future research that foreground lived experience within the home.

# 2. Related Work

## 2.1 Assistive mobility and lower-limb disabilities

Lower-limb disabilities encompass a range of mobility impairments resulting from paralysis, accidents, amputations, or congenital absence of one or both limbs, significantly affecting everyday movement and interaction with the built environment. Assistive technologies for mobility include devices such as manual wheelchairs, walkers, crutches, prosthetics, and their motorised variants. Under the Technology-Related Assistance of Individuals with Disabilities Act of 1988, assistive technology is defined as any product or system intended to increase, maintain, or improve the functional capabilities of individuals with disabilities (P.L. 100-407, 1988).

Mobility-related assistive technologies constitute the most widely used category of assistive devices, with an estimated 6.4 million users relying on them (Phillips & Zhao, 1993). Prior work differentiates between alternative devices, designed for individuals with total mobility impairment, and augmentative devices, which support users with residual mobility capacities (Martins et al., 2012). Research has shown that user requirements vary significantly depending on context. Studies comparing domestic and employment environments indicate that sturdiness, efficiency, and durability are prioritised in work settings, whereas domestic contexts emphasise simplicity, comfort, and ease of use (Biddiss & Chau, 2007).

## 2.2 Domestic space, spatial constraints, and abandonment

The domestic environment plays a critical role in shaping the everyday experiences of people with lower-limb disabilities. However, disability has historically been marginalised in housing and architectural research, with impairment rarely treated as an integral aspect of domestic habitation (Imrie, 2004). Architectural layouts, circulation paths, and spatial allocations are often designed around normative assumptions of able-bodied use, resulting in environments that fail to accommodate diverse bodily needs (Imrie & Hall, 2001; Heylighen & Strickfaden, 2012).

Empirical studies of domestic living with mobility impairments consistently identify spatial constraints such as narrow passages, thresholds, steps, inaccessible kitchens, and bathrooms as major barriers to independent movement (Imrie, 2004; Milner et al., 2001). Participants frequently describe homes as spaces that demand continuous adaptation, where everyday activities-such as making a bed, cooking, or bathing-become sites of friction. These experiences align with Leder's concept of the "dys-appearing body," in which bodily limitations become salient through persistent misalignment with designed environments (Leder, 1990).

Such domestic constraints are closely linked to the abandonment of assistive technologies. Abandonment is commonly defined as the non-use of a previously acquired or prescribed device (Phillips & Zhao, 1993). Research consistently reports high abandonment rates, with estimates suggesting that approximately one-third of assistive devices are eventually unused or stored away (Phillips & Zhao, 1993; Scherer, 2005). Key contributing factors include lack of user involvement in device selection, poor fit with domestic spaces, discomfort, stigma, and misalignment between professional goals and users' lived priorities (Federici et al., 2016; Scherer, 2005).

While assistive technologies are often framed by professionals in terms of physical functioning, users frequently evaluate them in relation to social participation, autonomy, and dignity-criteria that are insufficiently reflected in design and prescription processes (Scherer, 2002). These challenges are further intensified in underserved or low-resource contexts, where modifying domestic environments or accessing alternative devices is often financially or structurally infeasible (Imrie, 2004).

## 2.3 Participatory and speculative design in disability contexts

Participatory design approaches emphasise the inclusion of users as active contributors to the design process, particularly in contexts involving marginalised communities. Within disability research, participatory methods have been advocated as central to accountability, justice, and the legitimacy of design outcomes (Light et al., 2017). However, prior work in HCI and design research has critically noted that many disability technologies continue to be developed with limited involvement of disabled people themselves (Pullin, 2009).

Speculative design and related approaches-including design fiction, critical design, and futures-oriented methods-have been employed to explore alternative possibilities beyond existing technological paradigms (Dunne & Raby, 2013). These approaches aim not to predict future outcomes, but to open spaces for reflection, critique, and discussion. Design fiction, in particular, situates speculative futures within recognisable everyday contexts through narrative and artefacts (Bleecker, 2009; Dunne & Raby, 2013).

Despite increasing interest in participatory and speculative methods, their application within disability-related design remains limited. Prior work has highlighted that speculative practices are often framed from the perspective of designers or researchers rather than as co-constructed processes with disabled participants (Höök et al., 2015). Moreover, the use of speculative methods to examine everyday domestic mobility and assistive device use is underexplored.

## 2.4 Gaps in existing work

Across the literature, several gaps remain evident. First, although abandonment of assistive technologies is well documented, there is limited work examining how abandonment is shaped by domestic spatial conditions and everyday adaptations. Second, while participatory approaches are widely endorsed, insufficient attention has been paid to how different participatory media and prompts influence what participants are able or willing to articulate (Light et al., 2017). Third, speculative design methods are often applied without sustained engagement with lived domestic contexts, limiting their relevance to everyday assistive mobility practices (Dunne & Raby, 2013).

These gaps point to the need for participatory and co-speculative approaches that foreground domestic experiences of mobility, attend to limits of articulation, and treat dialogue itself as a meaningful research outcome. Addressing these gaps is essential for understanding persistent abandonment of assistive mobility devices and for advancing design research practices beyond solution-centric paradigms.

# 3. Methods

This study is situated within a constructivist qualitative research paradigm, which recognises knowledge as co-constructed through participants' lived experiences and interpretations (Creswell, 2013). The research adopts a participatory speculative approach to examine how people with lower-limb disabilities articulate challenges related to assistive mobility within domestic contexts. Rather than treating participants as sources of requirements, the study positions them as contributors whose reflections, critiques, and silences shape the inquiry. The methodological design combined semi-structured interviews, domestic friction mapping, and bespoke booklet-based participatory speculative workshops(Desjardins et al., 2019). These methods were organised as a connected sequence intended to progressively build context, trust, and modes of engagement. The research process unfolded across several phases: positional framing and consent approval; participant introduction and ice-breaking; semi-structured interviews; identification of domestic frictions; bespoke booklet workshop sessions; and synthesis with practitioners and professionals working in assistive mobility and care.

Participatory design was employed to foreground participants' lived experiences and to acknowledge that expertise about everyday mobility resides with those who navigate domestic environments daily (Sanders & Stappers, 2008; Light et al., 2017). Speculative design was used as a complementary approach to challenge assumptions embedded in existing assistive technologies and to open space for reflection on alternative possibilities beyond current products and infrastructures (Dunne & Raby, 2013). Together, these approaches aimed to shift the researcher's role from evaluator to facilitator, allowing participants to guide the narrative through critique, reflection, and selective engagement.

Data collection took place across multiple sites in and around New Delhi, selected to engage participants from commonly accessed healthcare and rehabilitation facilities. In total, 22 participants with lower-limb mobility impairments took part in the study. Data sources included audio recordings from interviews and workshops, annotated bespoke booklets, participant sketches, photographs of domestic environments, and researcher field notes. Semi-structured interviews were used to gather background information on participants' mobility histories, assistive device use, and daily routines. Domestic friction mapping supported the identification of challenges within home environments. The bespoke booklet workshops invited participants to respond to prompts addressing architectural constraints, existing assistive devices, and speculative future scenarios using writing, annotation, stickers, and sketches. All audio recordings were transcribed, and visual and material artefacts were digitised and catalogued. Data were analysed using reflexive thematic analysis, following Braun and Clarke (2006, 2021). Coding was conducted inductively across transcripts, booklet annotations, sketches, and images, with attention to recurring patterns, contrasts, and modes of articulation. Verbal, visual, and material contributions were treated as complementary sources of evidence rather than hierarchical forms of data.

The analysis was conducted in a reflexive manner, acknowledging the researcher's role in shaping interpretation across interviews, booklet responses, sketches, and field materials. Rather than treating themes as objectively discoverable, coding and theme development were understood as interpretive and shaped by ongoing engagement with the data, field notes, and the broader aims of the study. This approach was particularly important in understanding not only what participants articulated, but also where articulation stalled, shifted, or remained indirect.

This study was conducted as a pilot participatory workshop inquiry. Formal institutional ethics approval was not in place at the time of data collection. Permission to conduct the workshop and semi-structured interview at the Indian Spinal Injuries Centre (ISIC) New Delhi  and Community Health Centre (CHC) Dasna respectively,  was obtained from the host institution. Participants were informed about the purpose of the study, the voluntary nature of participation, and their right to withdraw at any time. Written informed consent was obtained before participation. All data used in the paper were anonymised, and identifiable participant images were blurred for dissemination.

# 4. Findings

## 4.1 Analysis Procedure

This study draws on data from two workshops conducted with 22 adult participants from age 17-55 y/o with lower limb disabilities and movement impairment. The workshop consisted of a semi structured interview followed by a Bespoke Booklet - based participatory speculative sessions. Data collected in both the sessions included audio recordings, transcripts, booklet annotations, sketches, photographs, video, notes. Audio recordings were selectively transcribed and visual and material artifacts were catalogued and photographed.

Data was analysed using reflexive thematic analysis. Coding was conducted across all data sources, attending to how participants articulated experiences, challenges, and critiques in response to different prompts and materials. Codes were iteratively reviewed and collected into themes based on recurring patterns and contrasts. The analysis treated verbal, visual, and material contributions as complementary sources of evidence, patterns traced across transcripts, booklet annotations, and sketches.

## 4.2 Forms of dialogue elicited

Participants were first engaged through one-on-one interactions using impromptu, semi-structured questions. Each interaction began with a brief introduction outlining the project, its motivation, and the overall research intent, as well as the participant's role in the study. Following this contextualisation, participants were asked about their disability, including its cause, and their history of assistive device use, such as the type of device used and the duration of use.

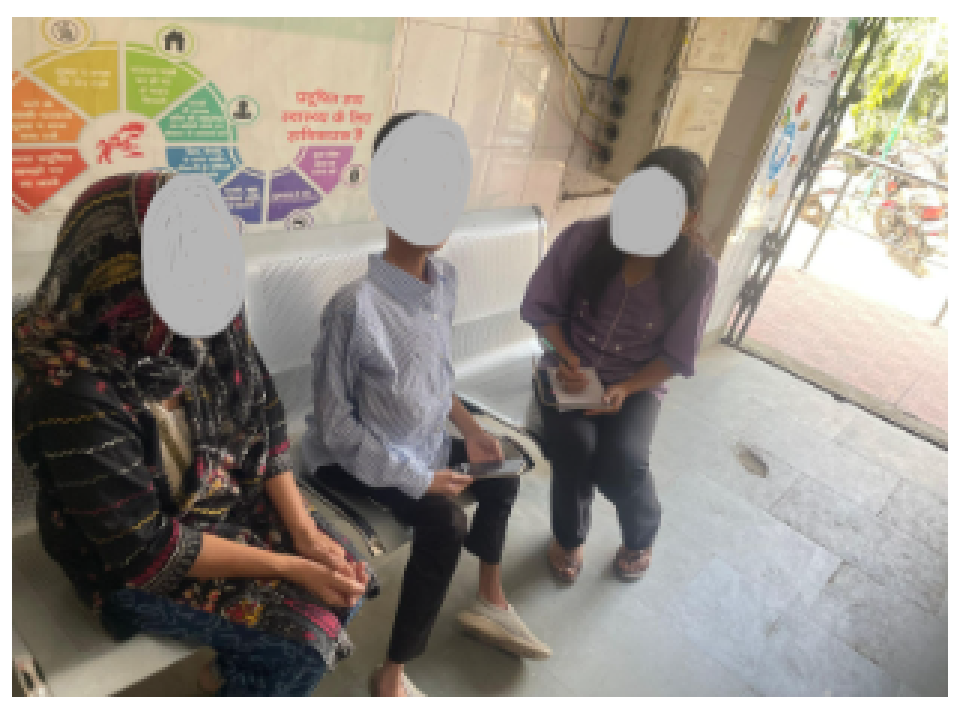
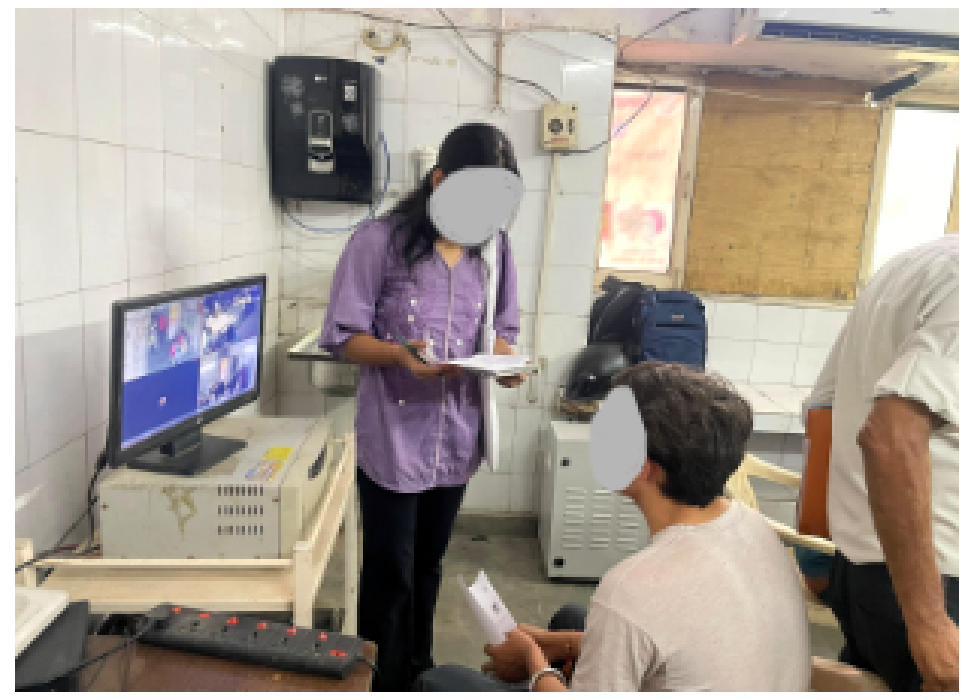
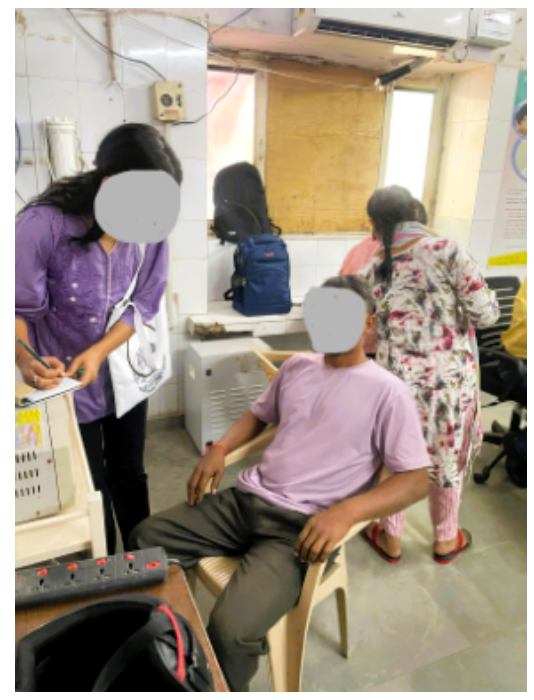

*Figure 1 Pictures taken during the semi-structured interview session at CHC Dasna, Ghaziabad.*

*Semi-structured interviews*

The semi-structured interviews were conducted at Community Health Centre (CHC) Dasna, Ghaziabad, a government hospital, during a routine disability camp (see Figure 1). Six participants aged between 17 and 40 years took part in the interviews. Participants represented a range of mobility impairments, including individuals who did not use assistive devices. The interviews consisted of five to six guiding questions, and participant responses shaped distinct forms of dialogue. An overview of the responses is presented in the table below.

*Table 1 Consolidated responses across semi-structured interviews*

| Q1: About (Age, Name, type of illness, and how long) | A1: Participants ranged in age from 17 to 40 years. Three participants had poliomyelitis and had been using wheelchairs since adulthood, with one reporting earlier use of crutches. Two participants had acquired lower-limb mobility impairments due to accidents; one had impairment in both legs and the other in one leg, and both used crutches. |
|---|---|
| Q2. Daily routine (Inside house mostly, assistance is required, if yes, by whom) | A2: Four participants reported requiring assistance for daily activities inside the home, particularly for using the bathroom and transferring between the bed and wheelchair when preparing to go out. These participants noted that wheelchairs were primarily used outside the home, while movement indoors relied on assistance from family members, such as parents or children. The remaining two participants described navigating indoor activities independently by using furniture for support. Most participants reported difficulty with bathing and toilet use, especially in the mornings; however, one participant highlighted the presence of a handrail in the bathroom as enabling greater independence for both bathing and toileting. |
| Q3. Challenges in Home | A3: Participants described multiple challenges related to movement within domestic spaces. One participant |
| | reported requiring regular assistance to access upper floors, including balconies and siblings' rooms. Four participants identified bathroom and kitchen thresholds as major obstacles when using crutches indoors, while a wheelchair user noted that the kitchen was largely avoided due to access difficulties, with bathroom thresholds remaining a concern. One participant highlighted difficulty caused by limited knee flexion when navigating raised pavements. Additionally, a participant with benign poliomyelitis who did not use assistive devices described the staircase leading to the terrace as a significant challenge. |
| Q4. Pain points in existing assistive devices used | A4: Participants consistently described existing assistive devices as bulky and unsuitable for use inside the home. Crutch users reported difficulty navigating narrow spaces, with crutches frequently getting stuck in corners when moving between rooms. Several participants expressed fear of slipping while bathing, despite the presence of rubber grips on crutches. One participant reported not using any assistive device indoors, instead relying on her hands and body for movement within the house and using a wheelchair only for hospital visits. Similarly, two crutch users stated that they avoided using assistive devices indoors, preferring to move independently or with support from walls and furniture. |
| Q5. Needs, wants and more | A5: Both female participants expressed a desire to carry out daily activities independently, particularly accessing and using the bathroom without assistance. The remaining participants predominantly raised concerns related to policy and government-subsidised assistive devices, describing them as bulky and unsuitable for use within the home. One participant with complete lower-body impairment noted that while advanced robotic wheelchairs could be helpful, they were prohibitively expensive and required substantial space for storage. Another participant described performing most household tasks independently and expressed pride in assisting his mother with small domestic activities despite having poliomyelitis. One female participant shared her interest in cooking and baking, noting that she rarely entered the kitchen because family members perceived it as too demanding for her; she expressed a wish to bake, influenced by online baking content she followed. |
| Q6. Emotions pertaining to themselves | A6: Five of the six participants reported experiencing pain, either when using assistive devices or when not |
| | using them. One participant who did not use any assistive device described joint pain when moving in and out of the house frequently. The remaining participants associated pain with specific aspects of device use, such as pressure from hard crutch cushions in the underarms, strain from gripping the central bar, or prolonged wheelchair use. These accounts primarily referred to pain experienced during movement outside the home, as most participants did not use assistive devices indoors. |

Following the interviews, audio recordings were transcribed and reviewed alongside field notes. Participants primarily expressed concerns related to mobility outside the home, while describing indoor mobility as something they had largely adapted to through informal strategies. Exceptions were noted in relation to daily routines such as transfers between bed and furniture, and access to kitchens and bathrooms. At the conclusion of each interview, participants were invited to voluntarily take part in a subsequent phase of the study. Two participants consented to continue-one woman and one man-and were asked to share photographs of challenging areas within their homes. Over the following ten days, a total of eleven images were collected, from which four were selected for analysis, representing the kitchen, bedroom, bathroom, and wheelchair storage area.

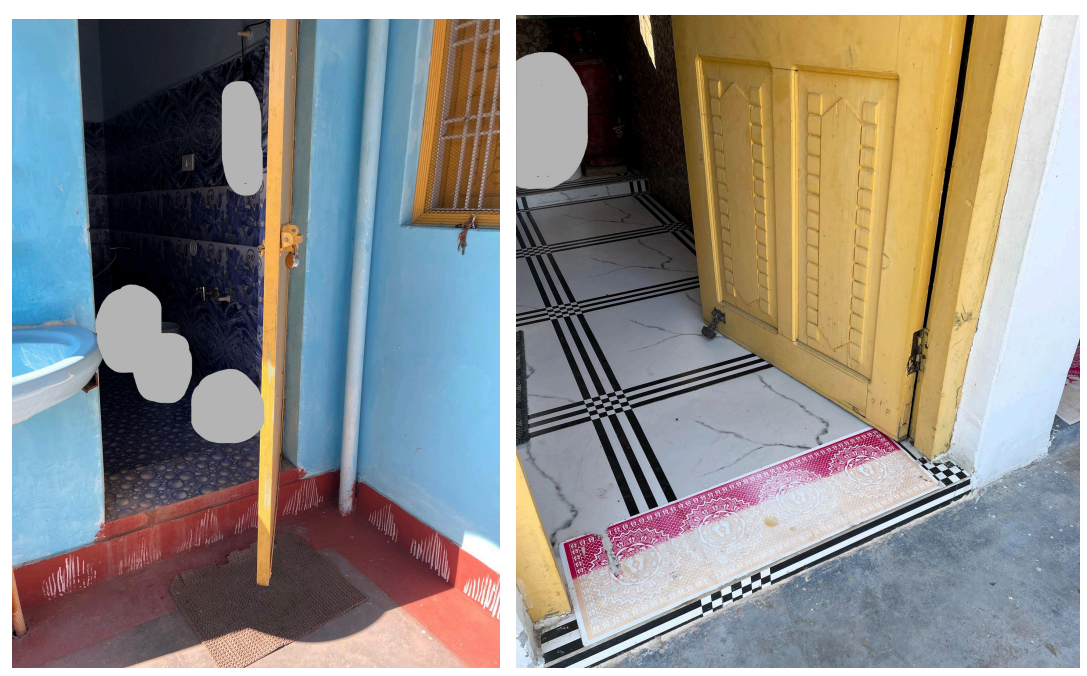

*Figure 2  Selected anonymised photographs of domestic interiors shared by participants, illustrating spatial constraints related to bathroom access, kitchen movement.*

The subsequent workshop phase was conducted across two locations at different times: with the two returning participants at the hospital, and with an additional sixteen participants at the ISIC centre in New Delhi.

*Bespoke booklet*

The bespoke booklet workshop was conducted at Indian Spinal Injuries Centre (ISIC), New Delhi, with sixteen participants (see Figure 3). Each participant was provided with a bilingual (Hindi-English) booklet, along with coloured pens, a pencil, and circular stickers. Following a brief introductory session outlining the purpose of the booklet and the research intent, participants were guided through the structure of the booklet and the ways in which they could engage with its pages. Participants were then given time to independently explore and respond to the booklet. Engagement with the booklet varied across participants. Twelve participants completed three to four of the seven pages, while one participant completed only two pages, stating that they had no comments on future technologies. Writing was generally minimal, with most participants contributing one to two sentences per page in Hindi. Only three participants engaged in drawing on more than one page; the remaining participants either drew on a single page or annotated directly on top of the provided images. During the session, participants twice asked whether they could add comments beyond the content already present in the booklet, subsequently recording additional opinions on the final page.

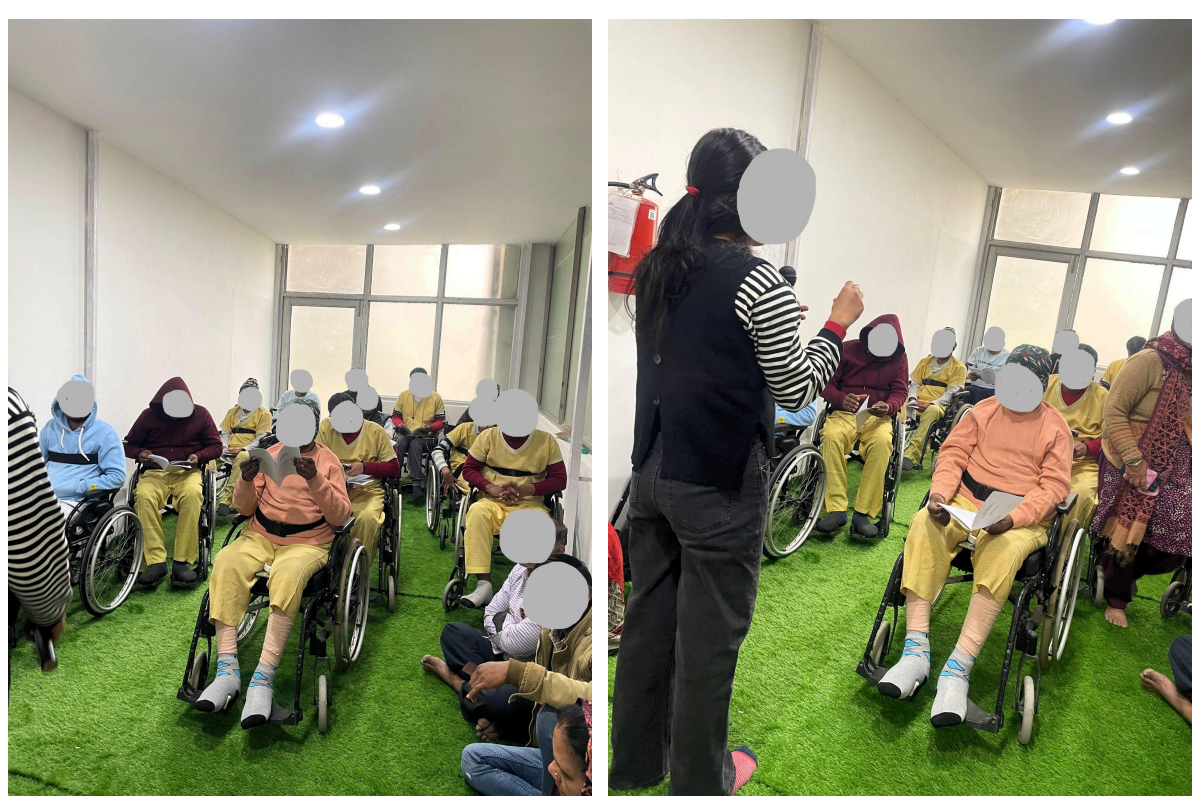

*Figure 3  Pictures taken during the speculative workshop session at ISIC, New Delhi.*

Distinct forms of dialogue emerged in response to different prompt categories (see Figure 3). Architectural and domestic-friction prompts elicited the highest level of engagement. Ten of the sixteen participants interacted actively with these pages by using stickers or sketches to indicate preferred counter heights and shelf positions within the home. In contrast, prompts focused on existing assistive devices generated limited engagement, with only four participants responding, primarily by noting the bulky nature of wheelchairs and the lack of appropriate storage space within domestic settings. Speculative future device prompts elicited the least engagement, with only two participants interacting with these pages. Responses in this category were predominantly evaluative, focusing on cost, availability, perceived complexity, and feasibility. While a DIY low-floor mobility concept prompted verbal questions regarding usability and adaptability, it did not result in written or drawn contributions.

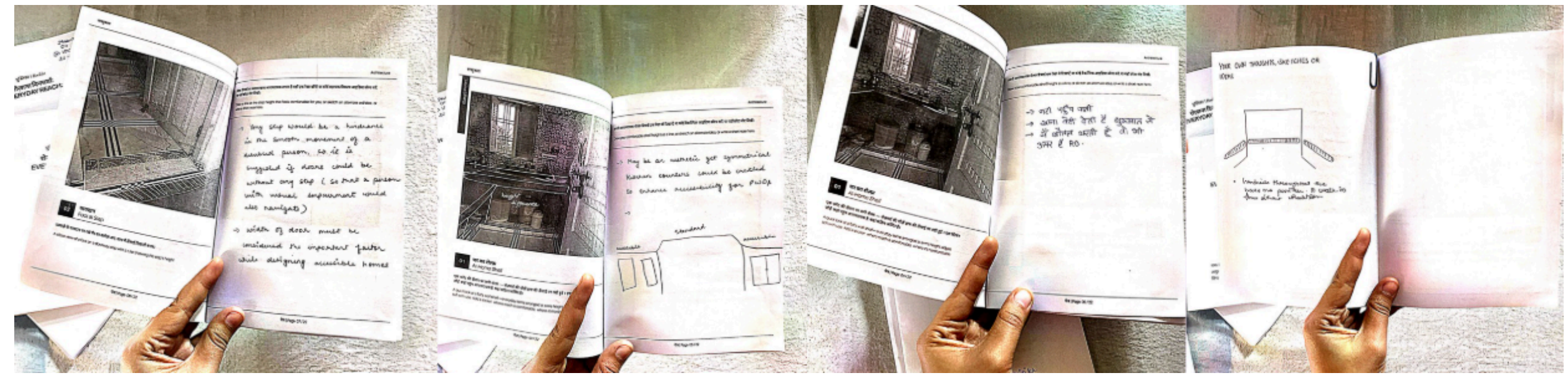

*Figure 4  Responses collected through sketches, notes and scribbles in the booklet.*

Across sessions, participants were more responsive to visual and contextual prompts than to writing-based speculative prompts. Engagement was largely reactive, taking the form of critique and comparison rather than generative ideation. Dialogue frequently shifted toward policy-oriented framing, with participants referring to limitations in government provision, subsidy structures, and access to innovative assistive technologies. Although domestic contexts were discussed, these were often referenced indirectly, with emphasis placed on institutional constraints rather than detailed descriptions of spatial or material conditions within the home.

*Table 2 Responses to Bespoke Booklet workshop session, showing variation in engagement across prompt types.*

| Prompt category (booklet pages) | Nature of prompt | Dominant mode of engagement | Level of engagement observed | Typical form of response |
|---|---|---|---|---|
| Architecture / Domestic layout (p. 4-9) | Visual photographs of shelves, steps, kitchen, bathrooms | Stickers, short annotations, occasional sketches | High | Identification of difficulty points; preferred heights; critiques of reach and stability |
| Assistive devices at home (p. 10-11) | Photograph of device storage and placement | Brief written comments | Low - Medium | Bulky, lack of storage space, unsuitability |
| Current assistive technologies (p. 12-14) | Images of existing market products | Evaluative written responses | Low | Cost, availability, perceived complexity |
| Speculative / fictional devices (p. 15-21) | Illustrated near-future and DIY concepts | Verbal critique; minimal writing | Very Low | Feasibility questions; rejection based on space, effort, cost. |

# 5. Discussion

This study set out to explore how people with lower-limb disabilities articulate domestic mobility challenges through participatory and speculative design methods. Rather than yielding concrete design proposals or device concepts, the findings provide insight into how dialogue emerges, where it stalls, and what this implies for assistive mobility research in domestic contexts.

## 5.1 Domestic mobility as a normalised condition

A key observation is that domestic mobility challenges were often described as managed rather than explicitly framed as design problems. Participants reported adapting to constrained household environments through informal strategies, family assistance, or avoidance of certain spaces. This normalisation of constraint helps explain why domestic challenges remain underrepresented in assistive technology research, despite their clear impact on everyday life. When limitations are treated as an expected condition of home life, they become difficult to articulate as sites for intervention, even within participatory settings.

This finding aligns with prior work on disability and domestic space, which shows how homes are frequently designed around able-bodied assumptions and how impaired bodies are required to adapt to these environments rather than the reverse (Imrie, 2004). The persistence of such adaptations suggests that assistive mobility research focused primarily on public accessibility risks overlooking where much of everyday life actually unfolds.

## 5.2 Policy-oriented discourse as an analytic signal

Participants frequently framed their concerns in relation to government policy, subsidy schemes, and the availability of assistive technologies, particularly when prompted to discuss improvements or future possibilities. Rather than treating this tendency as deflection, it is important to recognise policy-oriented discourse as an analytic outcome. These responses reflect how assistive mobility is experienced as a systemic issue, shaped by institutional provision and infrastructural constraints rather than individual choice alone.

This perspective complicates dominant narratives that position abandonment as a consequence of poor device design or user non-compliance. Prior studies have shown that abandonment is strongly linked to lack of user involvement in selection processes, limited choice, and misalignment between professional priorities and users' lived needs (Phillips & Zhao, 1993; Scherer, 2005). The findings of this study reinforce the importance of viewing abandonment within broader socio-technical and policy contexts.

## 5.3 Limits of speculative participation in domestic contexts

The study also highlights limits in how speculative methods function in disability-related research. While visual and contextual prompts supported active critique (see Figure 4), writing-based and future-oriented speculative prompts elicited minimal engagement. Participants were more comfortable responding to concrete representations of existing environments than imagining or proposing alternative futures. This suggests that speculative participation does not automatically translate into imaginative expression, particularly when participants' lived experiences are shaped by long-term constraint and dependency.

These observations resonate with critiques of speculative design that caution against abstract or designer-driven imaginaries that are insufficiently grounded in everyday practice (Dunne & Raby, 2013). In domestic assistive mobility contexts, speculation may require more embodied, situated, or performative methods to support articulation beyond critique and evaluation.

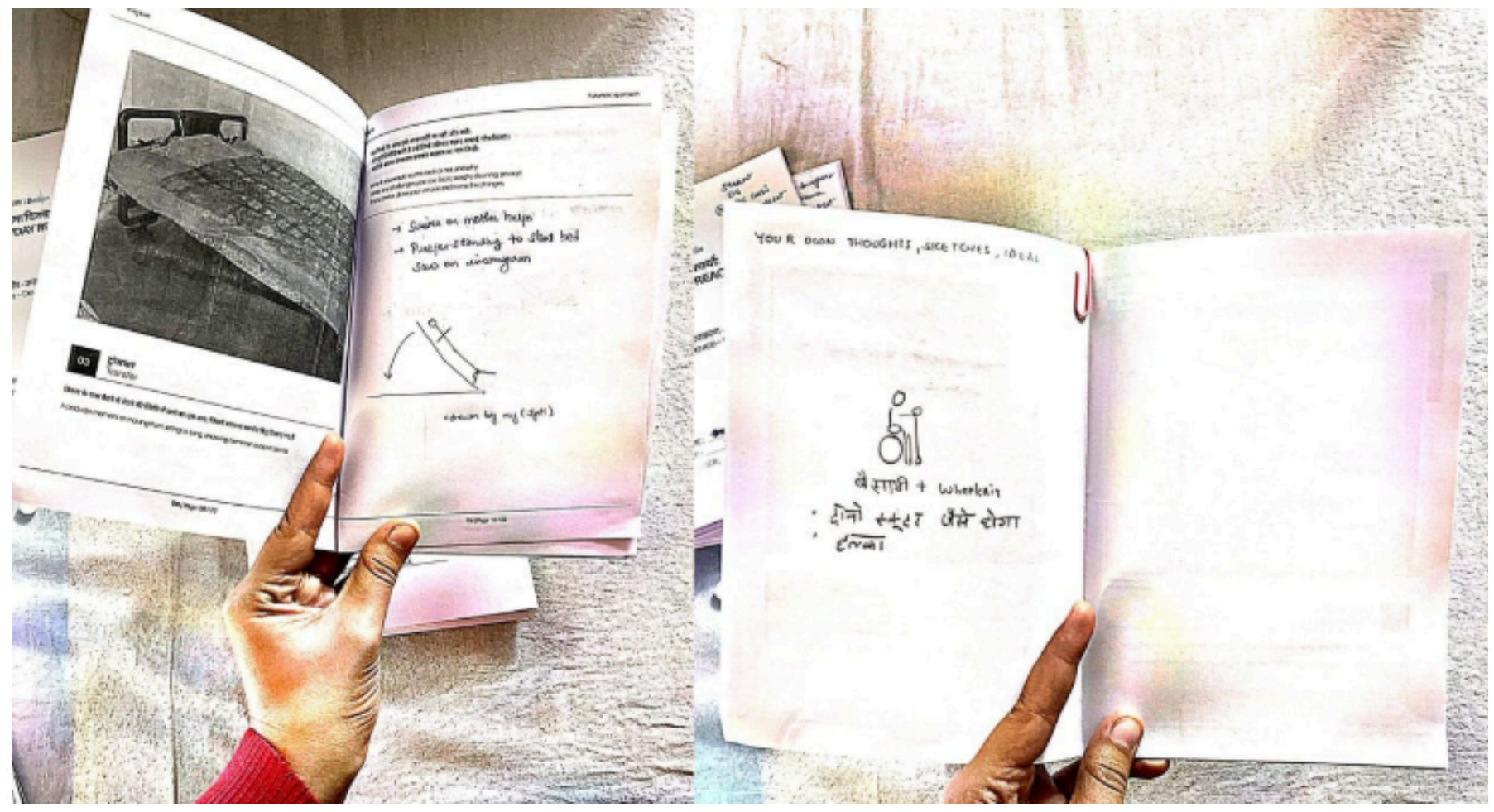

*Figure 5 Further responses and critique collected through sketches, notes and scribbles in the booklet.*

## 5.4 Methodological implications and future directions

The outcomes of this study point to the importance of method selection and sequencing. The bespoke booklet functioned effectively as a tool for eliciting critique and comparison, but was less effective in supporting generative ideation. Recognising such limits is crucial for advancing participatory design research in disability contexts, where articulation itself may be constrained by long-standing adaptation and social expectations (Light et al., 2017). Building on these insights, subsequent phases of this research will explore alternative participatory formats, including more embodied and performative methods such as but not limited to Magic Machine workshops (Andersen & Wakkary, 2019). Further approaches can be positioned as future work informed by the limits observed in the current study and are intended to lower abstraction, shift participants from respondents to co-participants, and support exploration of domestic mobility through action and interaction rather than reflection alone.

## 5.5 Reframing contribution

Taken together, the findings suggest that the value of participatory and speculative research in assistive mobility lies not solely in generating solutions, but in revealing how domestic mobility is experienced, articulated, and at times left unspoken (Phillips & Zhao, 1993; Scherer, 2005). By attending to what participants critique, avoid, or struggle to express, the study offers situated insights into the social, spatial, and institutional forces shaping everyday mobility within the home. Treating dialogue itself as a research outcome provides a way to better understand domestic adaptation, persistent abandonment, and the limits of existing assistive technologies, while also contributing methodological insight into how participatory and speculative methods operate in disability-related contexts.

# 6. Conclusion

This paper examined how participatory and speculative design methods can be used to initiate dialogue around domestic mobility challenges experienced by people with lower-limb disabilities. Through semi-structured interviews and bespoke booklet based workshops, the study explored how participants articulated everyday constraints, adaptations, and critiques of existing assistive technologies within home environments.

Rather than producing concrete design solutions, the study highlights the value of dialogue itself as an outcome. The findings show that domestic mobility challenges are often normalised and managed through informal strategies, making them difficult to surface through conventional needs-based approaches. While the bespoke booklet framework effectively supported critique of existing conditions and technologies, it was less effective in eliciting speculative or generative imaginaries, revealing important limits in how speculative participation operates in disability contexts.

By foregrounding these limits, this work contributes methodological insight to assistive technology and design research. It demonstrates how participatory speculative artefacts can expose not only unmet needs, but also points of hesitation, disengagement, and systemic constraints that shape everyday mobility and contribute to technology abandonment. The paper argues for greater attention to domestic contexts and to the forms of participation that enable people with disabilities to engage as co-contributors rather than respondents.

Overall, this study offers situated empirical and methodological insights for researchers and designers seeking to engage with assistive mobility beyond public infrastructures and solution-driven paradigms, while positioning domestic experience as a critical site for future inquiry.

**Acknowledgements:** The authors sincerely thanks Prof. Nishant Sharma, IDC, IIT Bombay, for his guidance and support throughout this work. The author is also grateful to Dr. Rimsha Siddiqui, Chief Operations Officer, ISIC New Delhi, for facilitating the workshop process, to Dr. Prerna Pant, Counselling Psychologist, for her support, and to Bhoomi Gupta, Masters student at IIT Roorkee for helping during the workshops. ChatGPT was used only for limited grammar correction and minor language refinement during manuscript revision.

About the Authors:

**Author 1** Jyoti Rautela is a M.Des. student at IIT Roorkee. A former accessory designer at Myntra and NIFT Bhubaneswar graduate, her interests include participatory, speculative, and critical design, HCI, affective computing, and inclusive design for adolescents/teens and older adults.

**Author 2** Abinash Kumar Swain is an Associate Professor in the Department of Mechanical and Industrial Engineering and Joint faculty in Department of Design at IIT Roorkee. His research interests include machine design, CAD, FEA, AI in design, interaction design, UI/UX, and product design.

**Author 3** Madhan Kumar Vasudevan is an Assistant Professor in the Department of Design at IIT Roorkee. His research spans human-computer interaction, haptics, VR/AR/XR, computational neuroscience, affective computing, machine learning, multisensory experience design, rehabilitation, healthcare design and engineering, and design for aging.